\documentclass[prb,twocolumn,showpacs]{revtex4}

\usepackage{graphicx}
\usepackage[caption=false,subrefformat=parens,labelformat=parens]{subfig}

\begin{document}

\hyphenpenalty=100000


\title{High-resolution magnetic penetration depth and inhomogeneities in locally noncentrosymmetric SrPtAs}

\author{J. F. Landaeta}

\author{S. V. Taylor}

\author{I. Bonalde}

\affiliation{Centro de F\'{\i}sica, Instituto Venezolano de
Investigaciones Cient\'{\i}ficas, Apartado 20632, Caracas
1020-A, Venezuela}

\author{C. Rojas}

\affiliation{Departamento de F\'{\i}sica, Facultad de Ciencias,
Universidad Central de Venezuela, Apartado 47586, Caracas 1041-A, Venezuela}

\author{Y. Nishikubo}

\author{K. Kudo}

\author{M. Nohara}

\affiliation{Department of Physics, Okayama University, Okayama 700-8530, Japan}

\date{02 February 2016}

\begin{abstract}
We present a magnetic penetration-depth study on polycrystalline and granular samples of SrPtAs, a pnictide superconductor with a hexagonal structure containing PtAs layers that individually break inversion symmetry (local noncentrosymmetry). Compact samples show a clear-cut $s$-wave-type BCS behavior, which we consider to be the intrinsic penetration depth of SrPtAs. Granular samples display a sample-dependent second diamagnetic drop, attributed to the intergrain coupling. Our experimental results point to a nodeless isotropic superconducting energy gap in SrPtAs, which puts strong constraints on the driven mechanism for superconductivity and the order parameter symmetry of this compound.
\end{abstract}


\maketitle

\section{introduction}

The recently discovered transition metal pnictide superconductor SrPtAs crystallizes in a hexagonal structure with Pt-As atoms in a single layer forming a honeycomb lattice,  \cite{nishikubo} differing from other pnictide superconductors that crystallize in a square lattice. The honeycomb 2D lattice of PtAs atoms locally breaks inversion symmetry, even though SrPtAs has a global center of inversion symmetry. The antisymmetric spin-orbit coupling (SOC) originating in Pt-As layers may allow a singlet-triplet mixing \cite{youn2011} that could lead to significant consequences in the superconducting properties of SrPtAs. Thus far, this mixing has manifested only in materials with a global lack of inversion symmetry.\cite{bauersigrist}

SOC at the Pt sites and locally broken inversion symmetry in the PtAs layers influence the electronic structure and the energy bands of SrPtAs.\cite{youn2012} Parity mixing is a consequence of the two bands that appear due to the lifting of the twofold spin degeneracy by the SOC. SrPtAs has three energy bands that cross the Fermi level and due to the SOC each of these bands splits at the Brillouin zone boundary face $k_z=\pi/c$ along the symmetry lines $H-A-L$.\cite{youn2011,youn2012,shein,goryo} On the other hand, even without SOC on the $k_z=0$ plane there is a splitting of the bands along the lines $K-\Gamma-M$ due to symmetry reasons. This splitting is proportional to the interlayer coupling.\cite{youn2011,youn2012} Then, the interplay of SOC (intraband) and interband interactions may become important. SOC favors parity-mixing effects, while interband interactions would support a conventional spin-singlet or spin-triplet superconductivity or could even lean toward a rare scenario of multiband superconductivity. Band-structure calculations suggest that SOC is larger than interlayer coupling.\cite{youn2011,youn2012}

Regarding the symmetry of the superconducting phase, three possible stable states have been suggested according to the point group $D_{3d}$: $E_g$, $A_{1g}$, and $A_{2u}$.\cite{goryo,fischer,akbari} These states allow parity mixing, despite the overall parity conservation in SrPtAs. The $E_g$ has a dominant chiral $d$-wave component and the $A_{2u}$ possesses a prevailing $f$-wave component.

Although there have been predictions that interesting phenomena may occur in SrPtAs, only a few experimental works have been performed. The findings are contrasting. Muon-spin rotation/relaxation ($\mu$SR) measurements carried out in polycrystalline and powdered SrPtAs  point to a superconducting state that breaks time-reversal symmetry (TRS) and has no extended nodes in the gap function.\cite{biswas} The zero-temperature penetration depths of the two tested samples were different by 30$\%$. No reason was reported for such a disparity. This study suggests that $E_g$, with a dominant chiral $d+id$-wave component that breaks TRS, is the most likely pairing state for SrPtAs. TRS breaking lifts some of the degeneracies still present in the band structure and, therefore, makes the analysis of the superconducting pairing symmetry more difficult.

In $^{195}$Pt nuclear magnetic resonance (NMR) and $^{75}$As nuclear quadrupole resonance (NQR) measurements in polycrystalline SrPtAs the relaxation rate $1/T_1$ displays a smooth curvature without anomalies. A coherence peak below $T_c$ and an exponential decay as temperature decreases were observed.\cite{matano} These results were interpreted in connection with an $s$-wave pairing symmetry with a gap $2\Delta_0=3.85 k_BT_c$, a value relatively high compared to the BCS standard. More recently, spin-lattice relaxation rate $1/T_1$ of $^{75}$As was measured in two different polycrystalline samples of SrPtAs.\cite{bruckner} No Hebel-Slichter coherence peak was observed and the temperature dependence of $1/T_1T$ -showing a strong hump around 300 mK- was interpreted in terms of two-gap superconductivity. The two samples showed NQR lines with different linewidth and slight asymmetry, both unusual effects attributed to impurities in the report.

From the existing studies, the prospects for SrPtAs are indeed interesting. SrPtAs could become the first example of a chiral superconductor with a $d$-wave symmetry, in analogy to Sr$_2$RuO$_4$,which has a chiral $p$-wave superconducting state.\cite{maeno5} On the other hand, the possibility of a compound with broken TRS and two energy gaps is an exciting scenario in superconductivity. While multiband descriptions are usually applied to ferromagnetic and global noncentrosymmetric superconductors, there is only one compound widely accepted to be a two-gap superconductor: MgB$_2$.~\cite{nagamatsu} This is a nonmagnetic centrosymmetric superconductor that conserves time-reversal symmetry. Two-gap superconductivity has not been established in nonmagnetic compounds with broken TRS. Thus, from both an experimental and a theoretical point of view, SrPtAs is a potential superconductor with combined exotic physical properties: parity mixing, multiple energy gaps and broken time-reversal symmetry. It is then of major interest for the general superconductivity community to clear up the current experimental results and their interpretations in SrPtAs.

Here, we report on a long-time study on several polycrystalline and powdered samples of SrPtAs with the intention to shed light on the superconducting gap structure and pairing symmetry. We present high-resolution magnetic penetration depth measurements down to 40 mK, a temperature which is well below the low-temperature limit of $0.2 \,T_c$ ($T_c=2.45$ K). The magnetic penetration depth $\lambda(T)$ is a direct response of the Cooper pairs and is widely considered one of the most powerful probes for the superconducting energy gap structure. Even though penetration-depth measurements in SrPtAs were reported previously,  \cite{biswas,nagamatsu} the far higher resolution of our technique to directly measure the penetration depth should allow us to better resolve features such as the inflation point in the superfluid density reported in Ref.~\onlinecite{biswas}. Since the marked inconsistencies between the previous experimental results may be due to sample quality, we also carried out a thorough study of the morphology and chemical composition of the SrPtAs samples using scanning electron microscopy (SEM) and energy-dispersive x-ray spectroscopy (EDS).

SEM analysis indicates that the polycrystalline samples of SrPtAs were heterogeneous, some of which were composed of aggregates of crystallites or grains in close contact (granular). EDS showed that there were no superconducting impurity phases in the samples. The penetration-depth results suggest that SrPtAs has a single isotropic energy gap.

\section{Experimental Details}

We measured several samples cut from a large polycrystalline SrPtAs piece grown as indicated elsewhere.\cite{nishikubo} To check on effects from inhomogeneities and intergranular coupling, one of the measured samples was grounded in a mortar, the resulting powder sedimented in alcohol and then cast in Stycast 2850. The SEM-EDS analyses were performed with a JEOL JSM-6400 scanning electron microscope and with a FEI Inspect F50 scanning electron microscope fitted with an Apollo X SDD x-ray microanalyzer.

Penetration-depth measurements were performed utilizing a 13.5 MHz tunnel diode oscillator.~\cite{mine10} The magnitude of the ac probing field was estimated to be less than 5 mOe, and the dc field at the sample was reduced to around 1 mOe. The deviation of the penetration depth from the lowest measured temperature, $\Delta\lambda(T)=\lambda(T)-\lambda(T_{min})$, was obtained up to $T \sim 0.99T_c$ from the change in the measured resonance frequency $\Delta f(T)$: $\Delta f(T) = G\Delta \lambda(T)$. Here $G$ is a constant factor that depends on the sample and coil geometries and that includes the demagnetizing factor of the sample. To within this calibration factor, $\Delta \lambda(T)$ is raw data. We estimated $G$ by measuring a sample of known behavior and of the same dimensions as the test sample,\cite{mine10} although in granular samples the effective geometry and hence the demagnetizing factor are somewhat unreliable. For the powder the susceptibility $\chi$ was related to $\lambda$ through $\label{chi} \chi=\frac{3}{2}\left \langle 1-\frac{3\lambda}{r}\mbox{coth}\frac{r}{\lambda} +\frac{3\lambda^2}{r^2} \right \rangle$. Here $r$ is the radius of a grain and $\langle \cdots \rangle$ denotes an average defined by $<x>\equiv \int x r^3 g(r)dr/\int r^3 g(r) dr$, $g(r)$ being the grain size distribution.

\section{Results and Discussion}

\begin{figure}%
	\centering
    \subfloat[]
	{\includegraphics[width=1.3in]{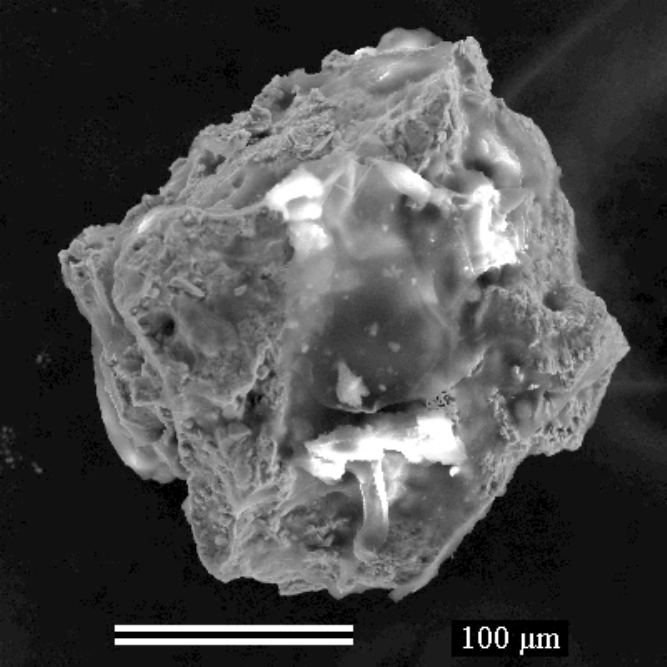}
		\label{fig:poli_A}}
	\subfloat[]
	{\includegraphics[width=1.3in]{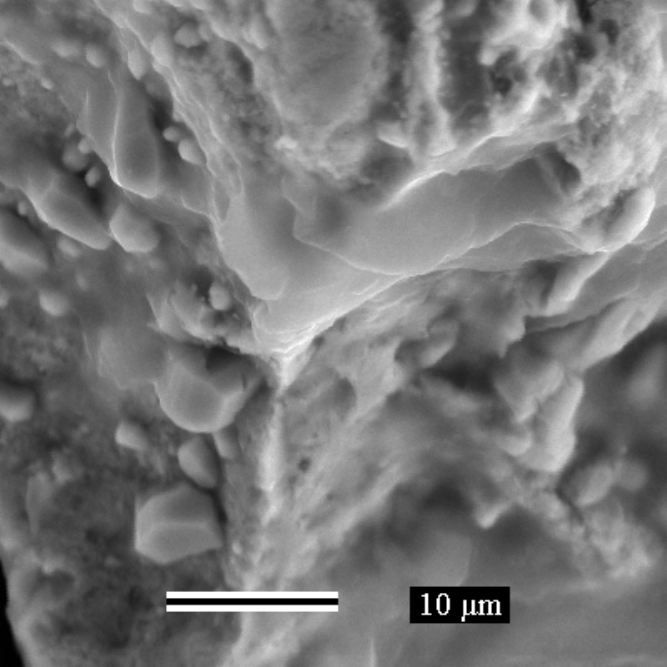}
		\label{fig:poli_A_zoom}
	}\\
	\subfloat[]
	{\includegraphics[width=1.3in]{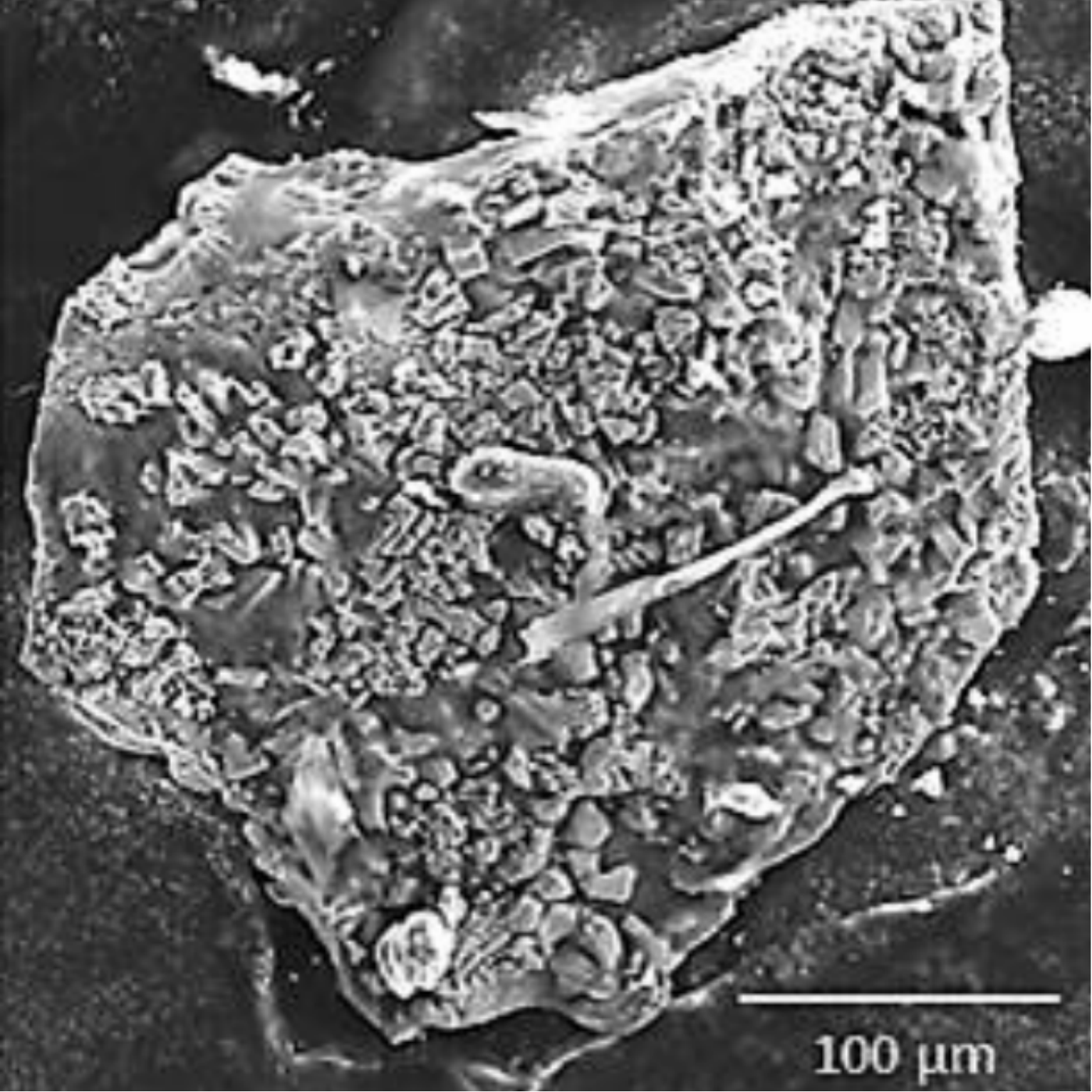}
		\label{fig:poli_B} }
	\subfloat[]
	{\includegraphics[width=1.3in]{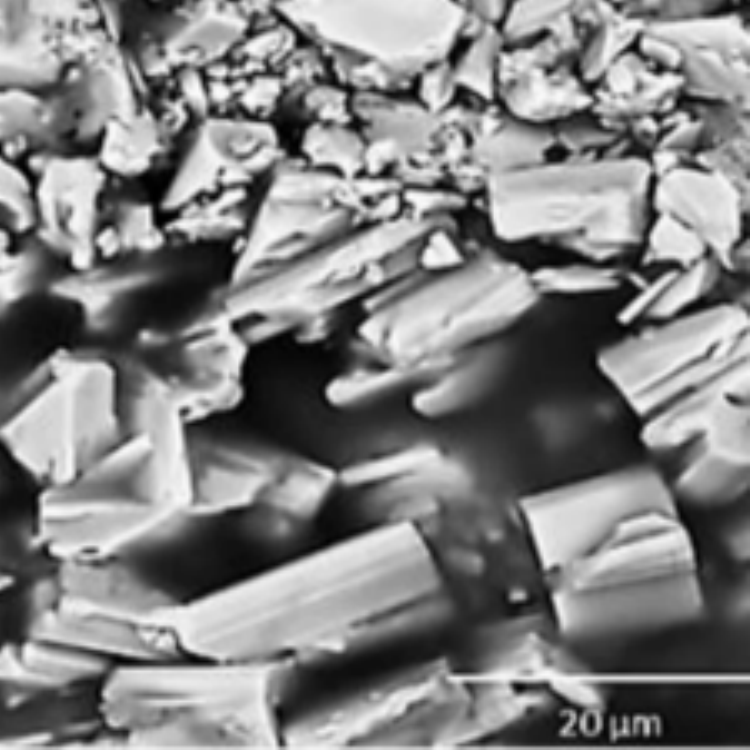}
		\label{fig:poli_B_zoom}
	}\\
	\subfloat[]
	{\includegraphics[width=1.3in]{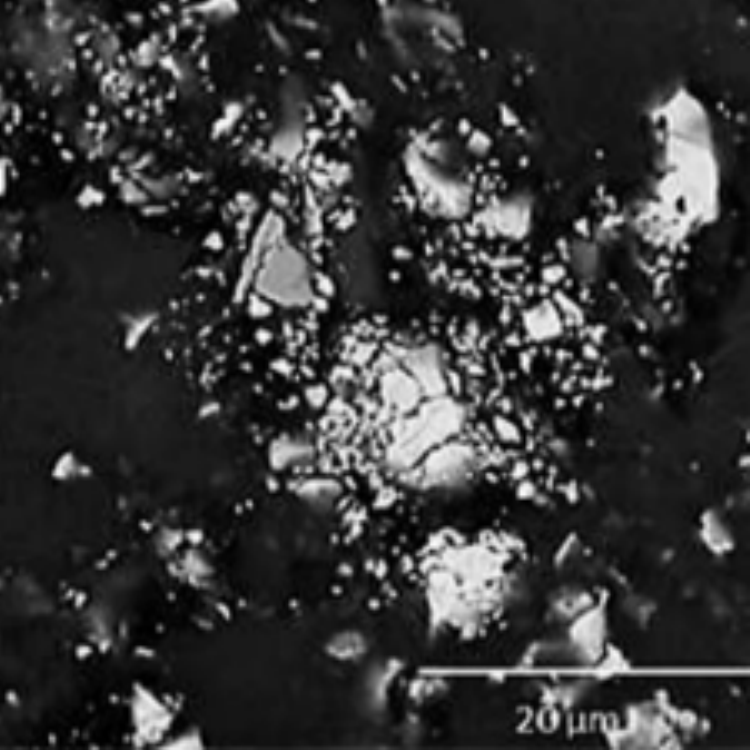}
		\label{fig:powder}}
	\caption{SEM images of (a) polycrystalline sample A, (b) a typical area of sample A, (c) polycrystalline sample B, (d) a typical area of sample B, and (e) a typical zone of the powdered sample. Sample B is evidently granular, whereas sample A looks compact. Note that the magnification in (b) is larger than in (d).}\label{fig:SEM}
\end{figure}

We used SEM-EDS analysis to study the chemical composition and morphology of the samples. For the elemental composition EDS study As-Sr L as well as Pt M signals were employed, because they were the more intense and were close together in energy. Information about the sample composition was obtained within 1.5 $\mu$m from the external surface. The sensitivity limit of our detector was about 1\%, although the error in the estimation of the elements was around 3\%. Spectra obtained from 40 crystallites of two different samples indicate that around 74\% corresponded to SrPtAs and the remaining 26\% to several nonstoichiometric SrPtAs phases, such as Sr$_{0.6}$PtAs$_{1.3}$, Sr$_{1.2}$Pt$_{0.8}$As and Sr$_{1.5}$Pt$_{0.6}$As$_{0.8}$. None of these nonstoichiometric compounds become superconductive to our knowledge. Thus, superconducting impurity phases in our SrPtAs samples were either not present or present below the sensitivity limit of our EDS system.

Secondary or backscattered electron SEM images were used for the morphology analysis. We discuss four polycrystalline samples labeled A, B, C, and D. Figure~\subref*{fig:poli_A} displays a SEM image of sample A, which was structurally heterogeneous but not granular (see details in Fig.~\subref*{fig:poli_A_zoom}). The SEM images of samples B, C, and D indicate that they had similar granular structures. Fig.~\subref*{fig:poli_B} shows an image of sample B and Fig.~\subref*{fig:poli_B_zoom} displays in detail the granular nature of this sample. The grains were faceted, with the larger ones having an elongated shape. Considering the largest dimension, the grain size ranged from about 1 to 20 $\mu$m. Figure~\subref*{fig:powder} presents a SEM image of the powdered sample, with the same magnification as in Fig.~\subref*{fig:poli_B_zoom}, showing some granular clusters and zones of well-separated grains with an average size of about 0.7 $\mu$m.

\begin{figure}
    \centering
	\includegraphics[width=3in]{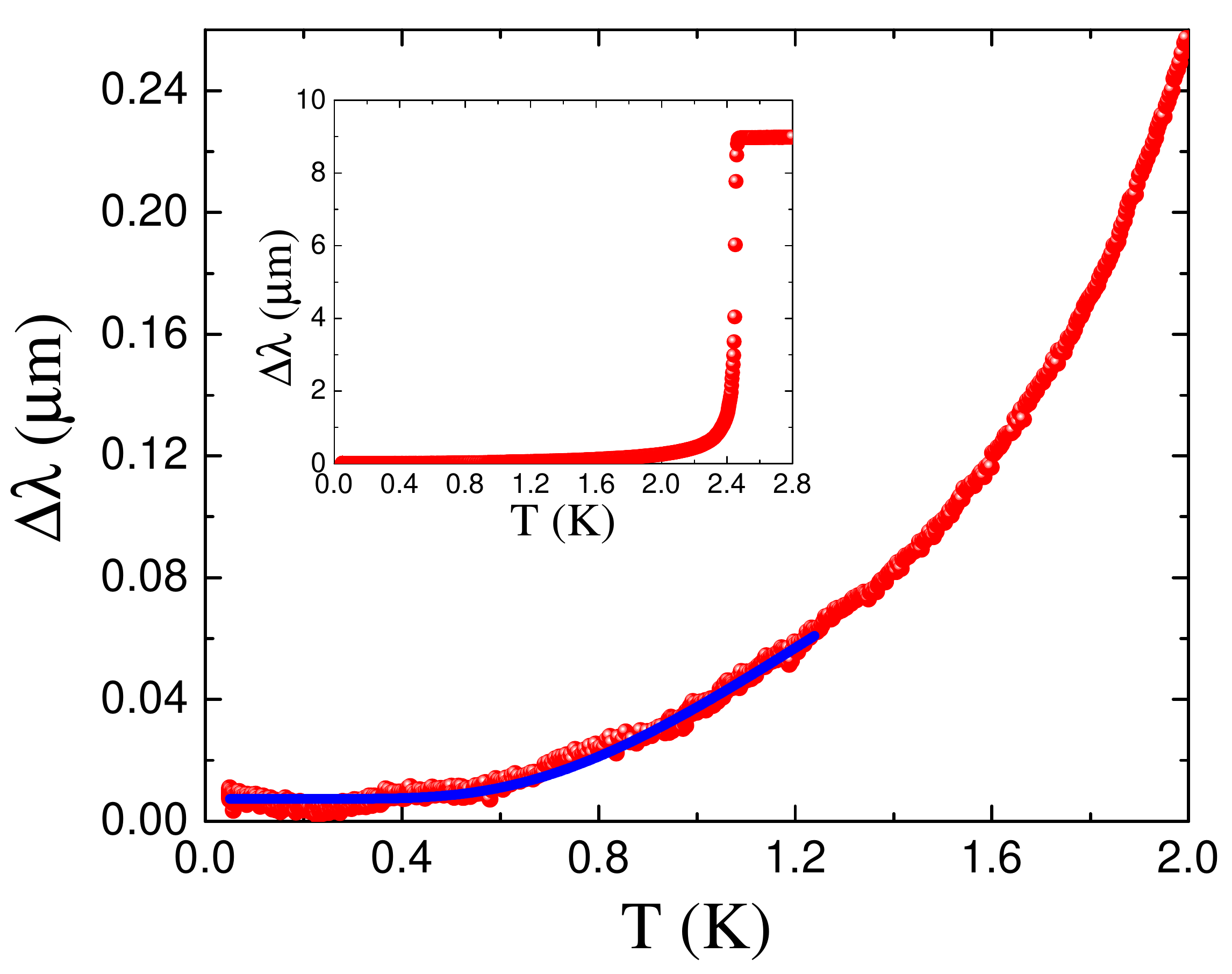}
	\caption{\label{fig:BCStype}{Magnetic penetration depth of polycrystalline SrPtAs. The main panel exhibits the low-temperature behavior of polycrystalline sample A. The solid blue line corresponds to the $s$-wave BCS model. The inset shows the behavior in the temperature region below $T_c$.}}
\end{figure}


The nongranular sample A, carefully and specifically selected to be as compact and less heterogeneous as possible, exhibited a clean $s$-wave-type penetration depth at low temperatures, as seen in the main panel of Fig.~\ref{fig:BCStype}.  Another compact sample (not discussed here) showed a similar low-temperature behavior. The inset of Fig.~\ref{fig:BCStype} displays the penetration depth of sample A in the whole temperature region below $T_c=2.45$ K. The penetration depth flattens below 0.4 K ($\sim 0.2T_c$), as expected for a superconductor with an isotropic energy gap.

At low temperatures $T<0.5T_c$, the data of sample A are very well fitted to the $s$-wave BCS model (blue line in the main panel of Fig.~\ref{fig:BCStype})
\begin{equation}
\label{swavebcs} \Delta\lambda(T) \propto \sqrt{\frac{\pi
\Delta_0}{2k_BT}} \exp(-\Delta_0/k_BT) \, ,
\end{equation}
with $\Delta_0=1.5 k_BT_c$, which is smaller than the
standard BCS value of $1.76k_BT_c$ and suggests that SrPtAs is in the weak-coupling regime. The value of $\Delta_0$ in the other compact sample not discussed here was similar ($\sim 1.55k_BT_c$).

\begin{figure}
    \centering
	\includegraphics[width=3in]{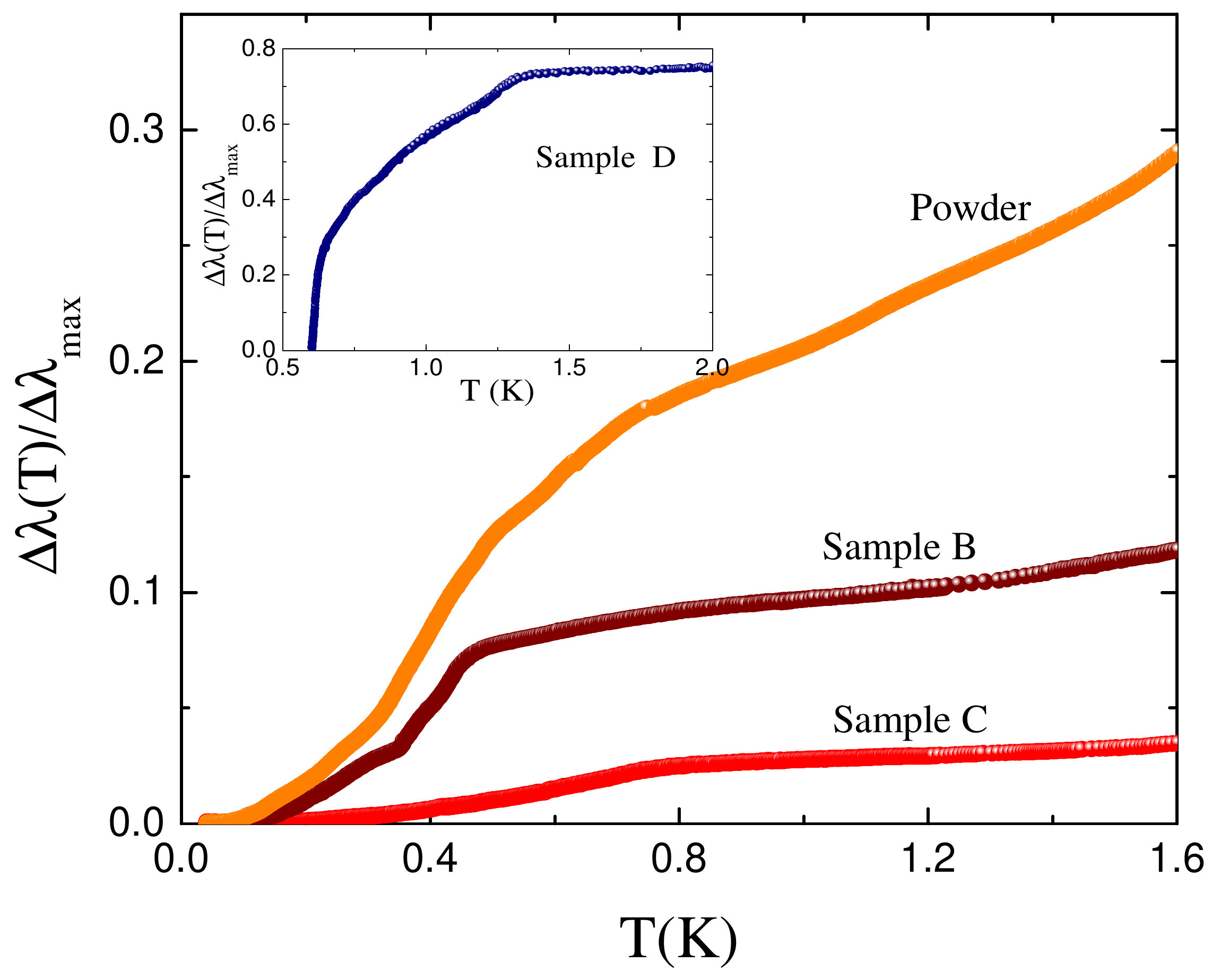}
	\caption{\label{fig:doubledrop}{Magnetic penetration depth of granular samples of SrPtAs. The main panel exhibits the low-temperature behavior of samples B, C and a powder, in which second diamagnetic drops are observed. The inset displays the low-temperature region for sample D that has a third drop. It is seen that the extra drops take place at different temperatures in the various samples, which suggests their extrinsic (nonsuperconducting) origin.}}
\end{figure}

Samples B and C as well as a powdered sample exhibited a second diamagnetic drop at different temperatures between 0.5 and 1 K (Fig.~\ref{fig:doubledrop}). Another granular sample (D) displayed even three drops between 0.5 and 1.5 K (see inset of Fig.~\ref{fig:doubledrop}). Samples B and D were subjected to slight stress with the intention of modifying the grain contacts. In granular superconducting materials the appearance of anomalies (extra drops or bumps) is usually caused by intergranular couplings, which are sample dependent.\cite{dxchen,goldfarb} Thus, the nonreproducibility and the sample dependance strongly evidence that the extra diamagnetic drops in samples B, C, and D are due to granular effects.

Our results appear to cast aside the existence of two-gap superconductivity in SrPtAs. A previous study of the NQR relaxation rate $1/T_1$ of SrPtAs shows a bump around 300 mK which was interpreted as two-gap superconductivity.\cite{bruckner} Since spin relaxation occurs internally in the grains, one may think that this effect would not be caused by granularity. It may be possible, though, that granularity leads to the slight asymmetry of the NQR lines and the difference in linewidth in samples A and B observed in Ref.~\onlinecite{bruckner}, as well as to the discrepancy in the zero-temperature penetration depth of samples A and B in Ref.~\onlinecite{biswas}.

The granular nature of most off-the-shelf polycrystalline samples of SrPtAs may lead to strong implications in some experimental studies. On the other hand, the EDS results may rule out impurity or secondary phases as the origin of the inconsistencies found in previous experiments.\cite{biswas,bruckner,matano}

To go even further with the discussion, we calculated the superfluid density of SrPtAs using the experimental data of sample A -which shows the intrinsic superconducting behavior. For this purpose, we used the zero-temperature penetration depth $\lambda(0) = 339$ nm.\cite{biswas} This value is consistent with $\lambda(0)=331$ nm calculated using $\xi(0) = 38.7$ nm (Ref.~\onlinecite{nishikubo}) and an estimated $B_c(0)= 19$ mT. With a Ginzburg-Landau parameter $\kappa\approx 9$, SrPtAs is a local superconductor. For such a superconductor the normalized superfluid density $\rho(T)= \left (n_s(T)/n \right )=\lambda^2(0)/\lambda^2(T)$, where $n$ is the total density, is given by

\begin{equation}
\label{supdens} \rho_s(T) = \left\langle 1 + 2  \int_{\Delta}^{\infty}  \left(\frac{D(E)}{D_0} \frac{\partial f(E)}{\partial E} \right) d E \right\rangle_{S.F}  \,.
\end{equation}

\noindent Here, $f$ is the Fermi function and $D_0$ is the density of states at the Fermi level. The total energy $E(T)=\sqrt{\epsilon^2 + \Delta^2(T)}$, and $\epsilon$ is the single-particle energy measured from the Fermi surface. We use here the standard weak-coupling gap interpolation formula $\Delta(T)=\Delta(0) \textrm{tanh}\left(\frac{\pi k_B T_c}{\Delta(0)} \sqrt{a (T_c/T - 1)}\right)$, where $\Delta(0)$ is the zero-temperature energy gap and $a$ is a constant related to the specific-heat jump at the superconducting transition and to the gap geometry.

\begin{figure}
\centering
\scalebox{0.35}{\includegraphics{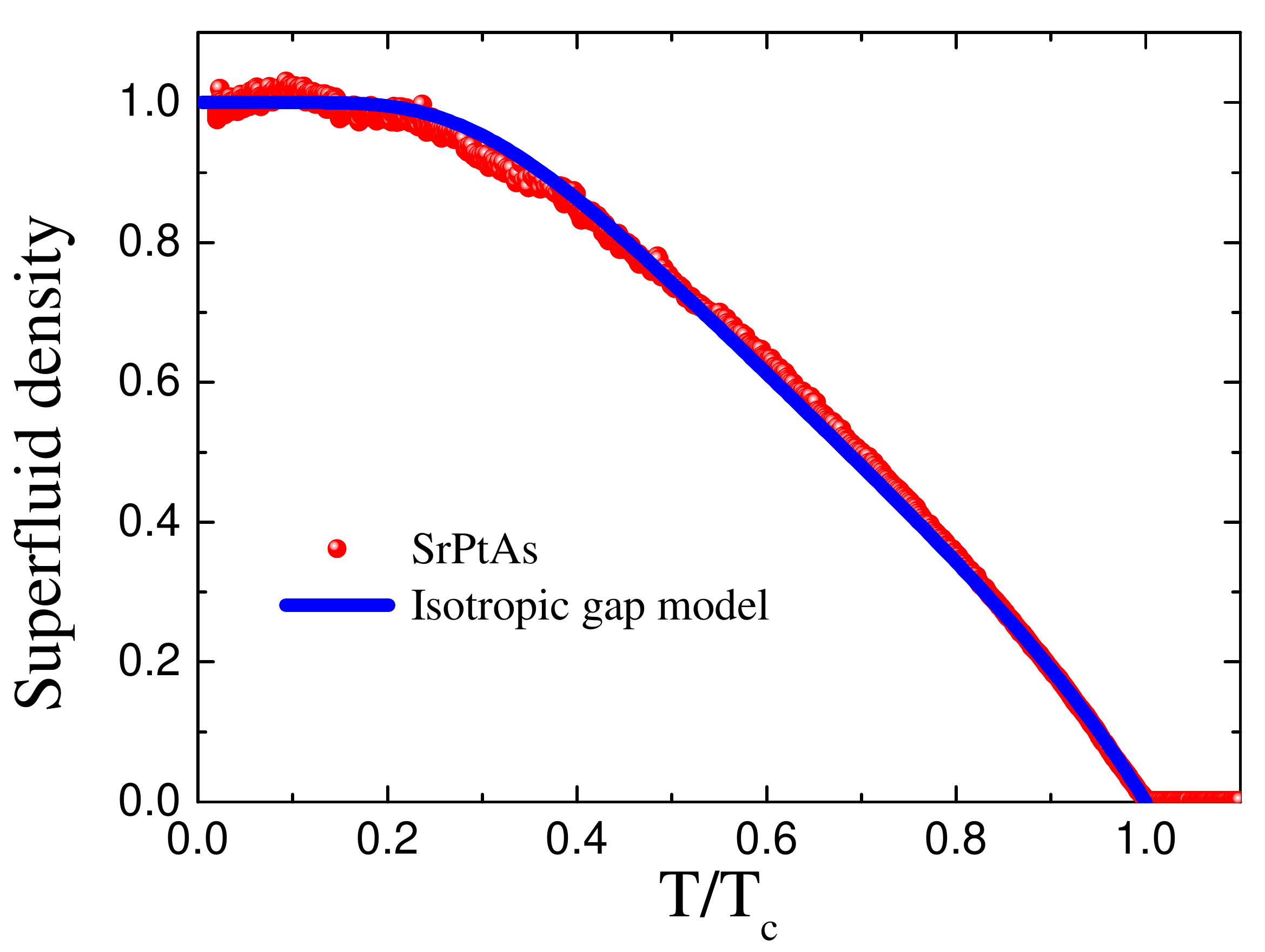}}
\caption{\label{fig:sup_density}{(Color online) Superfluid density of SrPtAs as a function of temperature compared to a BCS model with an isotropic energy gap.}}
\end{figure}

In Fig.~\ref{fig:sup_density} we compare the superfluid density of SrPtAs with an isotropic energy-gap BCS model (solid line).  The agreement is remarkable. This comparison yields the superconducting parameters $\Delta_0/k_BT_c=1.5$ and $a = 1.05$. These values are within the range of weak-coupling superconductivity. The gap value is smaller than $1.93k_BT_c$ obtained from $1/T_1$ measurements,\cite{matano} which were also interpreted in terms of an $s$-wave model.

The flat behavior at temperatures $T < 0.06 T_c$ in both the penetration depth and the superfluid density provides unambiguous evidence of the absence of nodes in the energy gap structure of SrPtAs, in agreement with some previous experimental works.~\cite{biswas,matano}

Band-structure calculations indicate that superconductivity mainly occurs in the pockets around $K$ and $K'$ points of the Brillouin zone. These pockets contribute most (about 70$\%$) to the total density of states (DOS) and have the largest superconducting gaps. The inner pockets around $\Gamma$, with the remaining 30$\%$ of the DOS, show very small gaps.\cite{fischer,wang} Regarding the order parameter symmetry, the $E_g$ state has a chiral ($d+id$)-wave symmetry with a nodeless and highly anisotropic structure on the pockets around $K$ and $K'$ points.\cite{goryo,fischer} For this state, the small pockets around $\Gamma$ are isotropic. Our results would be consistent with superconductivity mainly occurring in the small pockets and not in the ones around $K$ and $K'$, contrary to what is expected. Quantitatively, this is in line with the small gap value of $1.5k_BT_c$ found from our measurements. The $A_{1g}$, which has a dominant $f$-wave component, has a nodeless anisotropic gap on the pockets around the $K$ and $K'$ points and has nodes on the inner small pockets.\cite{fischer,wang} Thus, the dominant $f$-wave state seems to be incompatible with an isotropic gap. The $A_{2u}$ state with line nodes is fully discarded.

An isotropic energy gap for SrPtAs is a striking result, unexpected from a theoretical perspective. Most of the band-structure studies have indicated that SOC is larger than the competing interlayer interaction.\cite{youn2011,youn2012,wang,goryo,fischer} In global noncentrosymmetric superconductors an isotropic gap appears in general when the SOC strength is smaller than the superconducting gap (the other relevant energy scale).\cite{bauersigrist} An isotropic gap in SrPtAs, as found here and in Ref.\onlinecite{matano}, would somewhat imply that in this compound SOC could be comparable in strength with interlayer coupling and that parity mixing may not be important. But even neglecting parity mixing, it is still expected in SrPtAs an exotic $d$- or $f$-wave superconductivity, neither of which may be compatible with an isotropic gap.

It seems clear that studies in single crystals of SrPtAs would be beneficial to the understanding of its superconducting phase.

\section{Conclusions}

In summary, we performed magnetic penetration-depth measurements and SEM-EDS microanalyses in polycrystalline and powdered SrPtAs. In nongranular samples the penetration depth and the superfluid density display an $s$-wave-type behavior, suggesting that SrPtAs has an isotropic energy gap. This is not fully consistent with what is expected from band-structure analyses. The proposed symmetry states with a dominant $d$- or $f$-wave component are not fully compatible with an isotropic energy gap.

The extra diamagnetic drops observed in the penetration depth of granular samples are attributed to granular effects. Our results rule out the existence of two-gap superconductivity in SrPtAs.

\begin{acknowledgments}
We deeply appreciate conversations with M. Sigrist. We also thank G. Gonz\'{a}lez and M. Peralta at IVIC for their assistance with the electron microscopy study. This work was supported by IVIC Project No. 441, Venezuela.
\end{acknowledgments}


\end{document}